\documentclass[10pt, twocolumn, conference]{IEEEtran} %
\usepackage{subfigure, epsfig, color}
\usepackage{amsmath}
\usepackage{amsthm}
\usepackage{amssymb}
\usepackage{multirow}
\usepackage{booktabs}

\usepackage{mathtools}
\begin{document}

\title{Two-Stage Learning for Uplink Channel Estimation in One-Bit Massive MIMO}
\author{{
    Eren Balevi and
    Jeffrey G. Andrews}\\
    \IEEEauthorblockA{
		Department of Electrical and Computer Engineering \\
    The University of Texas at Austin, TX 78712, USA\\
    Email: erenbalevi@utexas.edu, jandrews@ece.utexas.edu}				
}

\maketitle 
\normalsize
\begin{abstract}
We develop a two-stage deep learning pipeline architecture to estimate the uplink massive MIMO channel with one-bit ADCs. This deep learning pipeline is composed of two separate generative deep learning models. The first one is a supervised learning model and designed to compensate for the quantization loss. The second one is an unsupervised learning model and optimized for denoising. Our results show that the proposed deep learning-based channel estimator can significantly outperform other state-of-the-art channel estimators for one-bit quantized massive MIMO systems. In particular, our design provides 5-10 dB gain in channel estimation error. Furthermore, it requires a reasonable amount of pilots, on the order of 20 per coherence time interval.
\end{abstract}


\section{Introduction}
Massive multiple-input multiple-output (MIMO) is a key technology to increase data throughput by allowing many users to concurrently use the same spectrum spatially \cite{Marzetta10}. This is achieved by eliminating interference among users thanks to the substantial degrees of freedom that comes from a large number of antennas. More specifically, this inter-user interference is mitigated by aligning beams properly for each user. This beam alignment conventionally requires good enough channel state information (CSI). Hence, acquiring CSI is of great importance in massive MIMO communication systems. 

Channel estimation has always been a challenge for communication systems due to the endless goal of reducing computational complexity and the number of pilot tones. This is particularly true for massive MIMO, because of the high signal dimension. Hence, massive MIMO channel estimation has to be performed with low complexity hardware and limited number of pilots, in which the former requirement is mainly to satisfy power consumption budget and the latter is for bandwidth efficiency. In this paper, these problems are handled for uplink multiuser massive MIMO. Although our main intent is for lower frequencies, it can be trivially applied for mmWave transmission.

Uplink massive MIMO nominally requires a correspondingly large number of analog-to-digital converters (ADCs) at the base station. This causes untenable power consumption and hardware complexity. There are many papers in the literature to alleviate these problems. These papers support employing low-resolution ADCs and analyze the impact of having a pair of one-bit quantization per each antenna, i.e., one for each real and imaginary component. More specifically, \cite{RisiLarsson14} shows the effects of one-bit quantization in terms of mutual information and symbol error rate for massive MIMO. A near maximum likelihood detector for one-bit uplink massive MIMO was designed in \cite{Choi16Heath}. Furthermore, \cite{MollenHeath17}, \cite{LiLiu17}, \cite{Wen16Ting} proposed some channel estimation methods against the detrimental impacts of one-bit quantization. 

In this paper, we propose to use deep learning methods for channel estimation in uplink massive MIMO with one-bit quantized received signals. This is motivated by the success of deep learning while coping with significant nonlinearities \cite{Goodfellow-et-al-2016}. There has been a growing interest to make use of deep learning in communication systems \cite{OShea17}, \cite{DornerBrink18}, \cite{Wen18Jin}, \cite{BalAndAntenna19} including channel estimation \cite{HeLi18}, \cite{yang2019deep}, \cite{Soltani18Sheikhzadeh}. The existing deep learning-based channel estimators rely on discriminative models. Note that a discriminative model is the one that simply maps the given input data or observations to their target values without using any a prior knowledge from these data. On the other hand, a generative model makes use of the a prior information while maximizing the likelihood. This motivation has recently inspired some deep generative model-based channel estimators \cite{BalAnd19}, \cite{BalAndDCE}, \cite{BalDosAndMassiveMIMO}.

The main contribution of this paper is to estimate the uplink channel in massive MIMO under the constraint of one-bit quantization by leveraging generative models. More specifically, a deep learning pipeline architecture is proposed for this problem. Our model is composed of two types of generative deep neural networks that were previously used for single antenna one-bit quantized OFDM channel estimation \cite{BalAnd19} and single antenna unquantized OFDM channel estimation \cite{BalAndDCE}. The proposed pipeline model works surprisingly well such that it can outperform state-of-the-art one-bit quantized channel estimators for uplink massive MIMO by 5-10 dB. Promisingly, this design only requires approximately $20$ pilots per coherence time interval.

This paper is organized as follows. The system model and problem statement are explained in Section \ref{Problem Statement and System Model}. The proposed architecture is given in Section \ref{Method}. We provide the numerical results and computational complexity analysis in Section \ref{Numerical Results}. The paper ends with concluding remarks in Section \ref{Conc}.

\section{System Model and Problem Statement} \label{Problem Statement and System Model}
In this paper, $K$ single antenna users send orthogonal OFDM symbols at the same time and on the same frequency band to a base station that has $M$ antennas. It is assumed that each OFDM symbol has $N_f$ subcarriers. With these settings, the received signal at the $m^{th}$ antenna in response to one transmitted OFDM symbol becomes
\begin{equation}\label{nth_OFDM_mth_ant}
\mathbf{y[m]} = \sum_{k=1}^K\mathbf{H_k[m]}\mathbf{F^{H}}\mathbf{x_k} + \mathbf{w[m]},
\end{equation}
where $\mathbf{H_k[m]}$ is an $N_f\times N_f$ circulant matrix, $\mathbf{F^{H}}$ is an $N_f\times N_f$ inverse discrete Fourier transform (IDFT) matrix, and the OFDM symbol $\mathbf{x_k}$ is an $N_f\times 1$ vector. The zero-mean Gaussian noise vector is represented by $\mathbf{w[m]}$. This received signal at each antenna port is quantized with a complex one-bit ADC. Thus, the real and imaginary part of the signal are quantized separately as
\begin{equation}\label{OFDM_quant}
\mathbf{r[m]}  = \mathbf{\mathcal{Q}(y[m])}
\end{equation}
where
\begin{equation} \label{quant}
\mathcal{Q}(\mathbf{y[m]})) = \frac{1}{\sqrt{2}}\text{sign}( \Re\{\mathbf{y[m]}\} )+\frac{j}{\sqrt{2}}\text{sign}( \Im\{\mathbf{y[m]}\} )
\end{equation} 
is applied element-wise along the vector $\mathbf{y[m]}$. Combining the received signal $\mathbf{r[m]}$ over all antennas yields an $N_f\times M$ matrix, which is
\begin{equation}\label{nth_OFDM}
\mathbf{R} = [\mathbf{r[1]}\ \mathbf{r[2]}\ \cdots\ \mathbf{r[M]}].
\end{equation}

In this paper, we assume that one coherence time is composed of $N$ OFDM symbols. The channel or $\mathbf{H_k[m]}$ remains constant during a coherence time interval 
and it periodically changes with coherence time interval. In this setup, the channel taps between the $k^{th}$ user and the $m^{th}$ antenna of the base station in the frequency domain is $\mathbf{\lambda_k[m]} = \text{diag}(\mathbf{\Lambda_k[m]})$, where
\begin{equation}
\mathbf{\Lambda_k[m]} = \mathbf{F}\mathbf{H_k[m]}\mathbf{F^{H}}.
\end{equation} 
Hence, the channel between the $k^{th}$ user and the base station for all the $M$ antennas becomes an $N_f\times M$ matrix
\begin{equation}
\mathbf{\Lambda_k} = [\text{diag}(\mathbf{\Lambda_k[1]})\ \cdots\ \text{diag}(\mathbf{\Lambda_k[M]})].
\end{equation}
Our problem is to estimate $\mathbf{\Lambda_k}$
from the received signal $\mathbf{R}$, which is defined in \eqref{nth_OFDM}, with some number of pilots smaller than $N$. These $\mathbf{\Lambda_k}$ matrices can be found separately for each user, because each user has orthogonal pilot sequences. 

\section{Deep Learning Pipeline for One-Bit Massive MIMO Channel Estimation} \label{Method}
\begin{figure} [!t] 
\centering 
\includegraphics [width=3.5in]{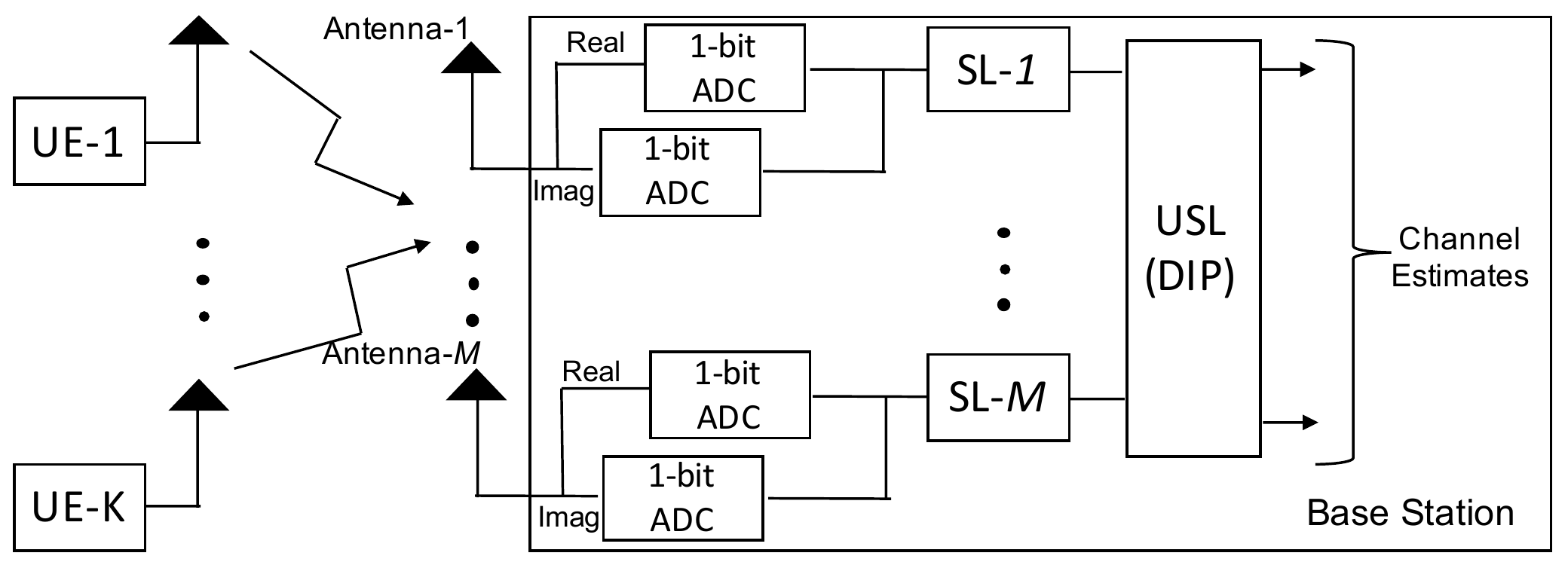}
\caption{One-bit quantized massive MIMO system in which $K$ users send their pilots to the base station with $M$ antennas. The received signal is first processed with $M$ supervised learning (SL) blocks, and then processed with one special unsupervised learning model, which is the deep image prior (DIP).} \label{fig:system_model}
\end{figure}

One-bit quantization leads to significant information loss at the receiver front-end. This considerably complicates channel estimation. We propose a pipeline deep learning architecture to reliably estimate the one-bit massive MIMO channel. Our architecture is composed of two separate deep learning models, each of which is specialized for different purposes. To be more precise, the first deep learning model, which is a generative supervised learning (SL) model, is designed for recovering the information loss due to the quantization. The second one, which is a generative unsupervised learning (USL) model, aims to denoise the received signal for channel estimation. The overall two-stage system model is depicted in Fig. \ref{fig:system_model}, in which the tasks from SL-1 to SL-$M$ involve the first stage and the USL task is the second stage.

The main reason behind the usage of generative models both for supervised and unsupervised learning is associated with the fact that generative models take into account the a priori knowledge as opposed to discriminative models. It is worth emphasizing that these priors are quite important for our problem, because one-bit channel estimation yields an ill-posed problem. This means that there is no hope to solve it without using priors. In particular, the quality of these priors determine the channel estimation error.

\subsection{Stage-1: Supervised Learning} 
\begin{figure} [!h] 
\centering 
\includegraphics [width=2.5in]{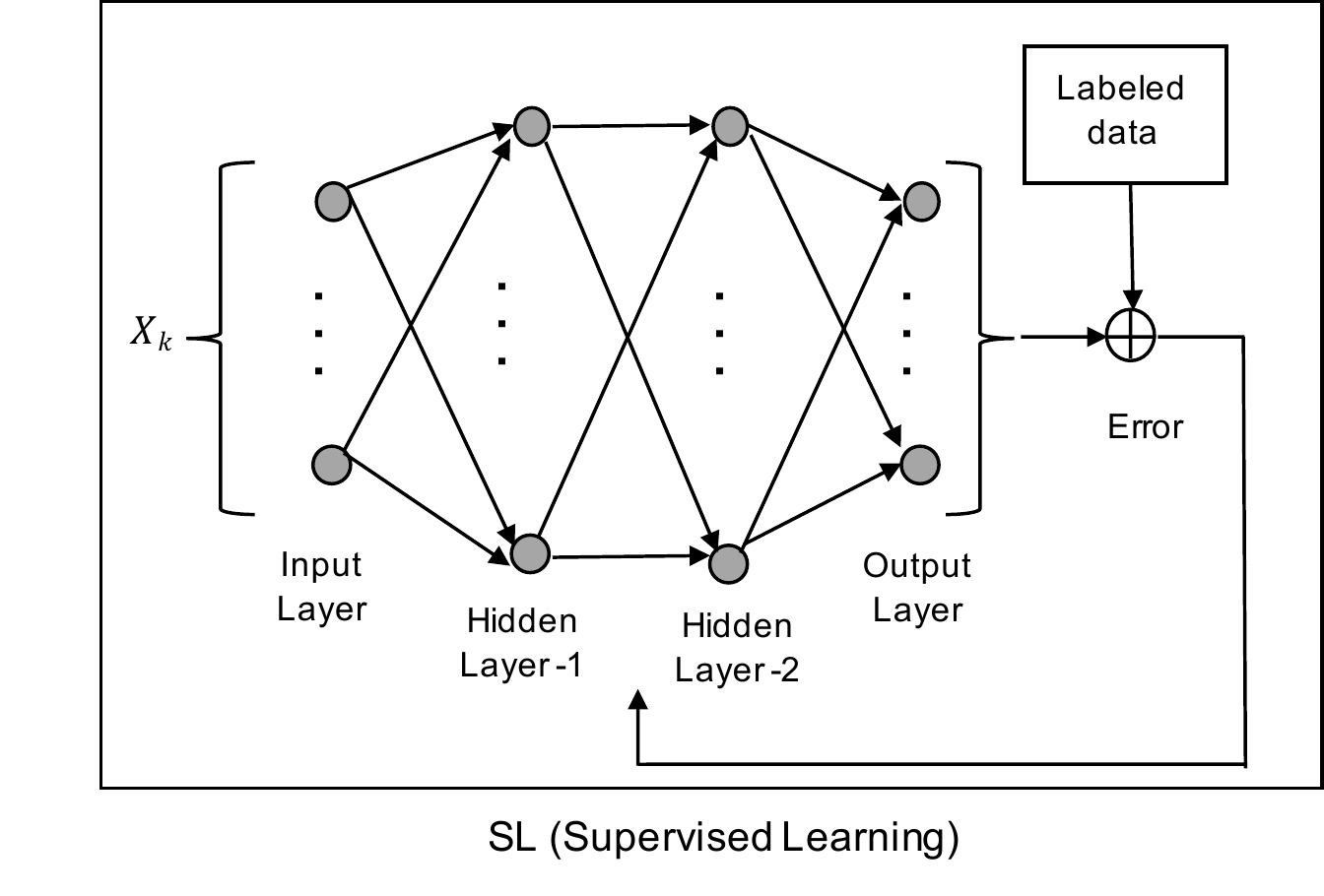}
\caption{The supervised deep learning architecture that is employed and trained separately for each antenna.} \label{fig:DL1}
\end{figure}

The received signal at each antenna port, which is given in \eqref{OFDM_quant}, is first processed with a separate supervised deep neural network. This means that $M$ deep neural networks are maintained in parallel. It is assumed that all these $M$ neural networks have the same architecture, which is a standard deep neural network that has two hidden layers as illustrated in Fig. \ref{fig:DL1}. It is worth emphasizing that the parameters of these $M$ deep neural network are not the same, because each one of them has its own labels and is trained individually according to their received signals and labels.

In supervised learning, the most critical thing is to define the labels intelligently to push the parameters of a neural neural towards the desired goal. Our recent work proposed generating a labeled data set on the fly for single antenna OFDM receivers using Bussgang's theorem \cite{BalAnd19}. Accordingly, the diagonals of the matrix $\mathbf{F}\mathbf{r[m]}\mathbf{x_k}^H$ are used as the labels. The theoretical ground for the selection of these labels is given in \cite{BalAnd19}. This labeling policy is also used in this one-bit massive MIMO channel estimation while training $M$ supervised deep neural networks. 

The layers, their types, sizes, activation functions and weights of the supervised deep neural network at each RF branch are summarized in Table \ref{tab:SLM}. State-of-the-art software libraries that implement neural networks do not support complex operations. Thus the real and imaginary part of the complex vectors are concatenated at each antenna to obtain a $2N_f\times1$ real vector. Without loss of generality, the dimension of the hidden layers is taken to be twice that of the input and output layer, giving $32N_f^2$ trainable parameters, which increases quadratically with the number of subcarriers. Notice that a single hidden layer can give the same performance with two hidden layers if it has a sufficient number of neurons due to the universal function approximation theorem of neural networks \cite{Goodfellow-et-al-2016}. However, this brings additional computational complexity, and hence having two hidden layers with reasonable number of neurons seems a good compromise. Rectified linear unit (ReLU) is used in the hidden layers as an activation function for fast convergence, and a linear activation function is utilized at the output layer. 
\begin{table} [!t] 
\renewcommand{\arraystretch}{1.3}
\caption{The supervised DNN architecture for channel estimation with one-bit ADCs}
\label{tab:SLM}
\centering
\begin{tabular}{c|c|c|c|c}
    \hline
       Layer & Type & Size & Activation & Weights\\
    \hline
    \hline
Input Layer & Pilot Symbols & $2N_f$ & - & $-$\\
    \hline
Hidden Layer-1 & Fully Connected & $4N_f$ & ReLU & $\boldsymbol{\Phi}_{1}$\\
    \hline
 Hidden Layer-2 & Fully Connected & $4N_f$ & ReLU & $\boldsymbol{\Phi}_{2}$\\
    \hline
 Output & Fully Connected & $2N_f$ & Linear & $\boldsymbol{\Phi}_{3}$\\
		\hline
\end{tabular}
\end{table}

We propose to use the aforementioned model as a regression task. Accordingly, the input layer takes the pilots $\mathbf{x_{k,p}}$ and produces the corresponding output $\mathbf{z_p[m]}$ for $p=1, \cdots, N_t$ where $N_t$ is the total number of pilots transmitted over the channel for one coherence interval, in which $N_t < N$. Notice that the pilots for the users are sent orthogonally, and hence can be treated individually. This $\mathbf{z_p[m]}$ for $m=1,2,\cdots,M$ can be written in terms of the trainable weights or network parameters (in matrix notation) and activation functions as
\begin{equation} 
\mathbf{z_p[m]}=\sigma_3(\boldsymbol{\Phi}_{3}\sigma_2(\boldsymbol{\Phi}_{2}\sigma_1(\boldsymbol{\Phi}_{1} \mathbf{x_{k,p}})))
\end{equation}
where $\boldsymbol{\Phi}_{i}$ and $\sigma_i$ are the network parameters and the activation function, respectively. The parameters are optimized according to the following cost function
\begin{equation} \label{cost_func}
J_m=\min_{\boldsymbol{\Phi}_{1}, \boldsymbol{\Phi}_{2}, \boldsymbol{\Phi}_{3}}\left|\left|{\mathbf{z_p[m]}}-\text{diag}(\mathbf{F}\mathbf{r[m]}\mathbf{x_{k,p}}^H)\right|\right|^2
\end{equation}
which are solved with gradient descent via the backpropagation algorithm.

The $M$ deep neural networks are trained separately to minimize the MSE between the outputs and the labeled data as given in (\ref{cost_func}). After training, for each supervised neural network we generate as many output samples as needed in response to random inputs within the same channel coherence interval, and take their average. The generated output samples for the random inputs do not cost anything other than some extra processing, because these inputs are not coming from the channel; rather they are generated randomly in the receiver. To be more precise, each trained deep neural network generates some output samples $\mathbf{z_i[m]}$ in response to the random inputs $\mathbf{x_{k,i}}$. In what follows, the output of each deep neural network is obtained as
\begin{equation} \label{est_chn_taps}
\mathbf{\hat{\lambda}_k[m]} = \frac{1}{N_g}\sum_{i=0}^{N_g-1}\mathbf{z_i[m]}
\end{equation}
where $N_g$ is the total number of arbitrarily generated output samples. There is no constraint to limit $N_g$ except the processing complexity, i.e., the $\mathbf{z_i[m]}$ does not consume any bandwidth. Each time the channel changes, the model must be retrained with $N_t$ pilots, and $N_g$ randomly generated samples. Note that there are many different types of generative model applications other than the generative adversarial networks (GANs) \cite{goodfellow2014generative}, and hence this is one type of generative models. 

\subsection{Stage-2: Unsupervised Learning} 
The main benefit of using a second deep learning model is to enhance the quality of channel estimates. In this regard, an unsupervised deep learning model is utilized. More precisely, the output of each supervised deep learning model that estimates the channel taps in the frequency domain given in \eqref{est_chn_taps} is replicated $N$ times in the time domain. This means that an $N_f\times N$ complex frequency-time grid is obtained for each user. Then, this frequency-time grid is stacked for all antennas for $m=\{1,\cdots,M\}$ in the spatial domain. This yields a $3$-dimensional $N_f\times N \times M$ signal for each user, which is denoted as $\mathbf{\Lambda_T}$. 

\begin{figure} [!t] 
\centering 
\includegraphics [width=3in]{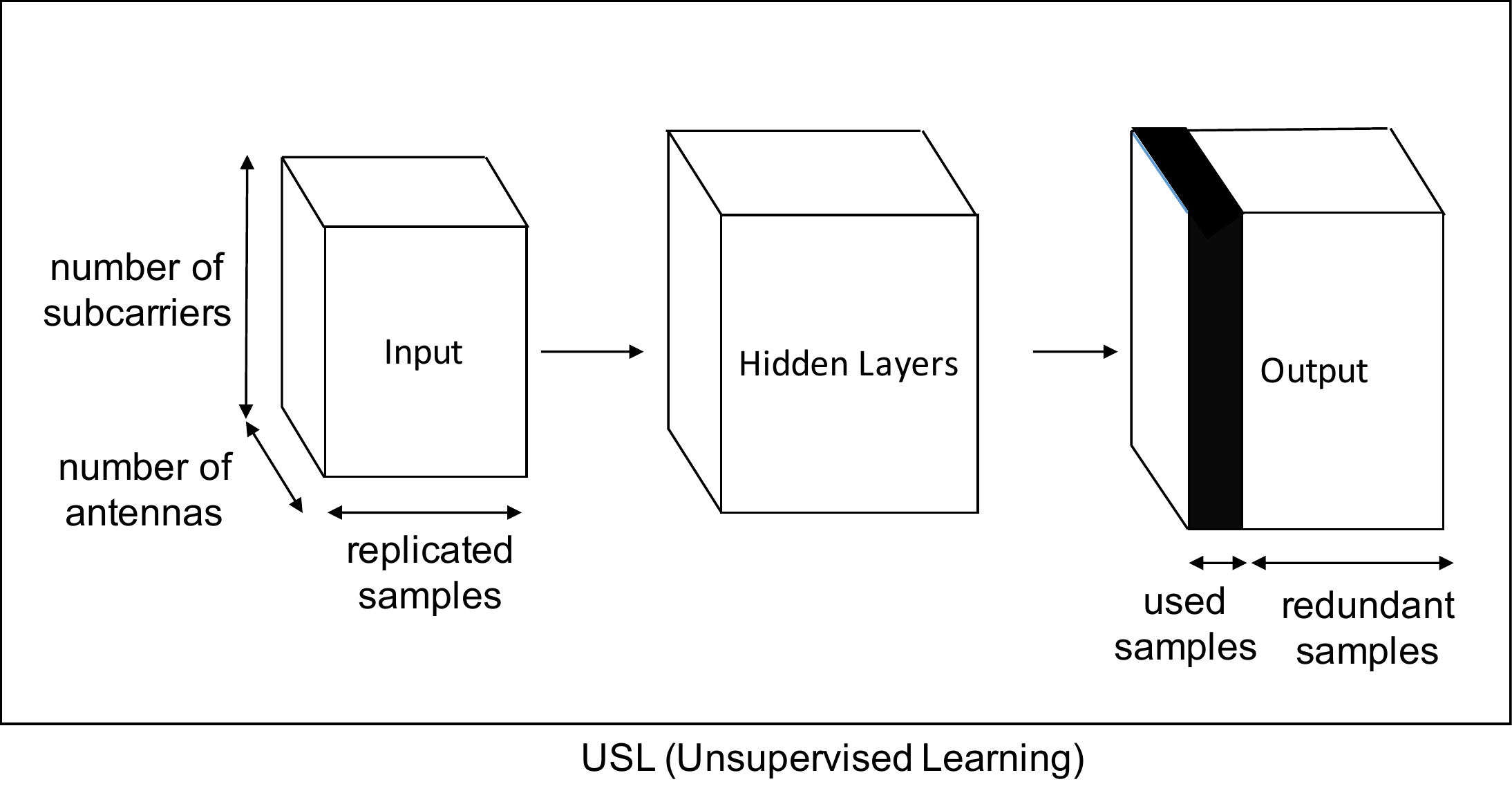}
\caption{The second unsupervised deep learning model whose output gives the channel estimates.} \label{fig:DL2}
\end{figure}

The unsupervised deep learning model processes this $3$-dimensional signal by fitting the parameters of its deep neural network. The overall unsupervised model that depicts the input, output and hidden layers is given in Fig. \ref{fig:DL2}. Here, the deep neural network targets to denoise the signal by fitting the signal more and the noise less. This is possible, because the signal has a structure, whereas the noise is unstructured, i.e., the noise is independent and identically distributed (i.i.d). In particular, the hidden layers are crafted so as to exploit the structure (or correlations) of the signal. The working principle of the model in Fig. \ref{fig:DL2} is to generate $\mathbf{\Lambda_T}$ by passing a randomly chosen input tensor $Z_0$ through hidden layers. Here, $Z_0$ is as an input filled with uniform noise and once it is randomly initialized, it is kept fixed. The weights of the hidden layers are also randomly initialized, but their weights are continuously updated via gradient descent.

\begin{figure*} [!t] 
\centering 
\includegraphics [width=6in]{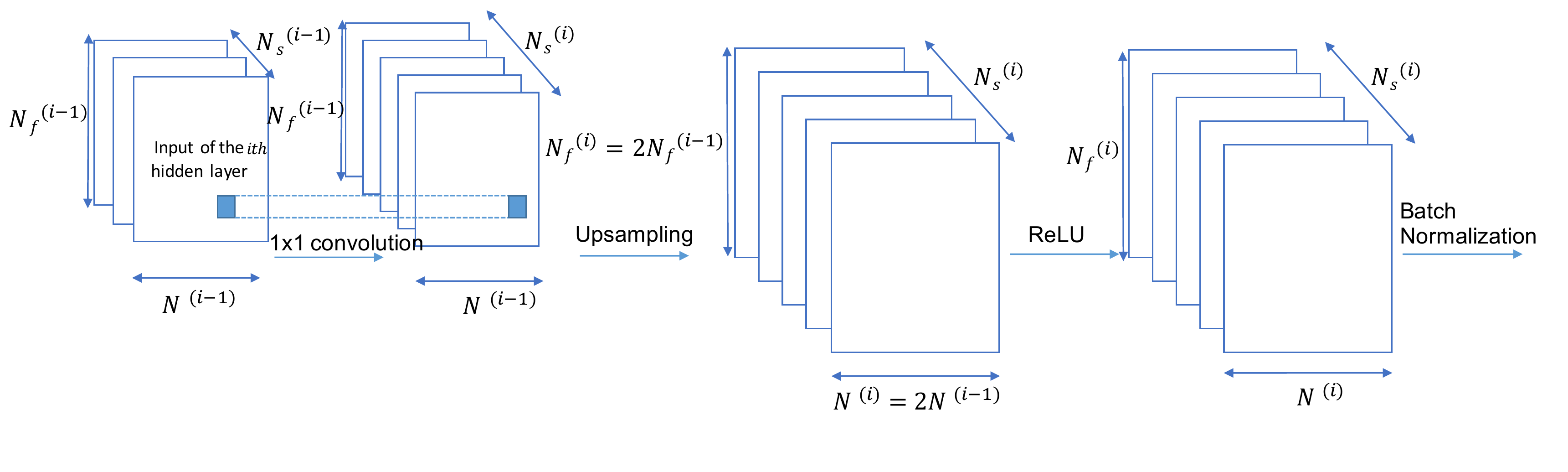}
\caption{The structure of the $i^{th}$ hidden layer, whose input dimension is $N_f^{(i-1)} \times N^{(i-1)} \times N_s^{(i-1)}$ and output dimension is $N_f^{(i)} \times N^{(i)}\times N_s^{(i)}$. Note that $N_f^{(i)} = 2N_f^{(i-1)}$ and $N^{(i)} = 2N^{(i-1)}$. The spatial dimensions $N_s^{(i-1)}$ and $N_s^{(i)}$ are the hyperparameters that are used by the $1\times1$ convolution operations.} \label{fig:model}
\end{figure*}
The key component in the aforementioned generative unsupervised model is the hidden layers. This structure of a hidden layer is portrayed in Fig. \ref{fig:model}. Each hidden layer is composed of four major components. These are: (i) a $1\times1$  convolution, (ii) an upsampler, (iii) a ReLU activation function, and (iv) a batch normalization. A $1\times1$ convolution means that each element in the time-frequency grid is processed with the same parameters through the spatial domain, which changes the dimension. There are $N_s^{(i)}$ different kernels, which are shared for each slot in the time-frequency axes. Hence, the spatial dimension becomes $N_s^{(i)}$. This can be equivalently considered as each vector in the time-frequency slot being multiplied with the same (shared) $N_s^{(i)} \times N_s^{(i-1)}$ matrix. In what follows, upsampling is performed to exploit the couplings among neighboring elements in the time and frequency grid. More precisely, the time-frequency signal is upsampled with a factor of $2$ via a bilinear transformation. Next, the ReLU activation function is used to make the model more expressive for nonlinearities. The last component of a hidden layer performs batch normalization for a batch size of $1$ to avoid vanishing gradients. All the hidden layers have the same structure except for the last hidden layer, which does not have an upsampler.

The mathematical representation of the aforementioned architecture is given next. Accordingly, the tensor $\mathbf{\Lambda_T}$ is parameterized for the $l+1$ layer as
\begin{equation}
\mathbf{\hat{\Lambda}_T} = f_{\theta_l}(f_{\theta_{l-1}}( \cdots f_{\theta_0}(Z_0)))
\end{equation}
where the input $Z_0$ has a dimension of $N_f^{(0)} \times N^{(0)} \times N_s^{(0)}$ in the frequency, time and spatial domain, respectively. These dimensions are determined according to the number of hidden layers and the output dimension, in which $N_s^{(l)}=M$,  $N_f^{(l)}=N_f$, $N^{(l)}=N$. The layers from $0$ to $l-1$ are counted as a hidden layer, and for $i=0,1,\cdots,l-2$
\begin{equation} \label{eq:Op1}
f_{\theta_{i}}=\text{BatchNorm}(\text{ReLU}(\text{Upsampler}(\theta_i\circledast Z_{i}))
\end{equation}
where $Z_i$ is the input of the $i^{th}$ hidden layer, $\theta_i$ are the parameters, and $\circledast$ represents the so-called ``convolution'' operator, which actually refers to cross-correlation in signal processing. More precisely, a $1\times1$ convolution is utilized as a cross-correlator, which means that the spatial vector for each element of the time-frequency grid is multiplied with the same shared parameter matrix to obtain the new spatial vector for the next hidden layer. The last hidden layer is
\begin{equation} \label{eq:Op2}
f_{\theta_{l-1}}=\text{BatchNorm}(\text{ReLU}(\theta_{l-1}\circledast Z_{l-1}),
\end{equation}
and the output layer is
\begin{equation}
f_{\theta_{l}}=\theta_{l}\circledast Z_{l}.
\end{equation}

All the parameters can be represented as 
\begin{equation}
\Theta = (\theta_{0}, \theta_{1}, \cdots, \theta_{l}),
\end{equation}
which are optimized according to the square of $l_2$-norm
\begin{equation}\label{obj_func}
\Theta^* = \arg \min_\Theta ||\mathbf{\Lambda_T}-\mathbf{\hat{\Lambda}_T}||_2^2.
\end{equation}
The output of the DNN for the optimized parameters is
\begin{equation}\label{outDNN}
\mathbf{\Lambda_T^*} = f_{\Theta^*}(Z_0),
\end{equation}
where the channel estimate $\mathbf{\hat{\Lambda}_k}$ corresponds to the 2-dimensional $N_f\times M$ signal at the first time index in \eqref{outDNN}. Notice that the spatial dimension of $Z_0$ is a hyper-parameter and its other dimensions are determined by the output signal and the number of layers, since at each layer the time and frequency dimensions are doubled. Furthermore, it is worth emphasizing that in the architecture spatial correlations in the signal are captured by the $1\times1$ convolution so as to decrease the number of parameters, and frequency and temporal correlations are exploited by upsampling. 

\section{Numerical Results and Complexity Analysis} \label{Numerical Results}
\begin{figure} [!h] 
\centering 
\includegraphics [width=3.5in]{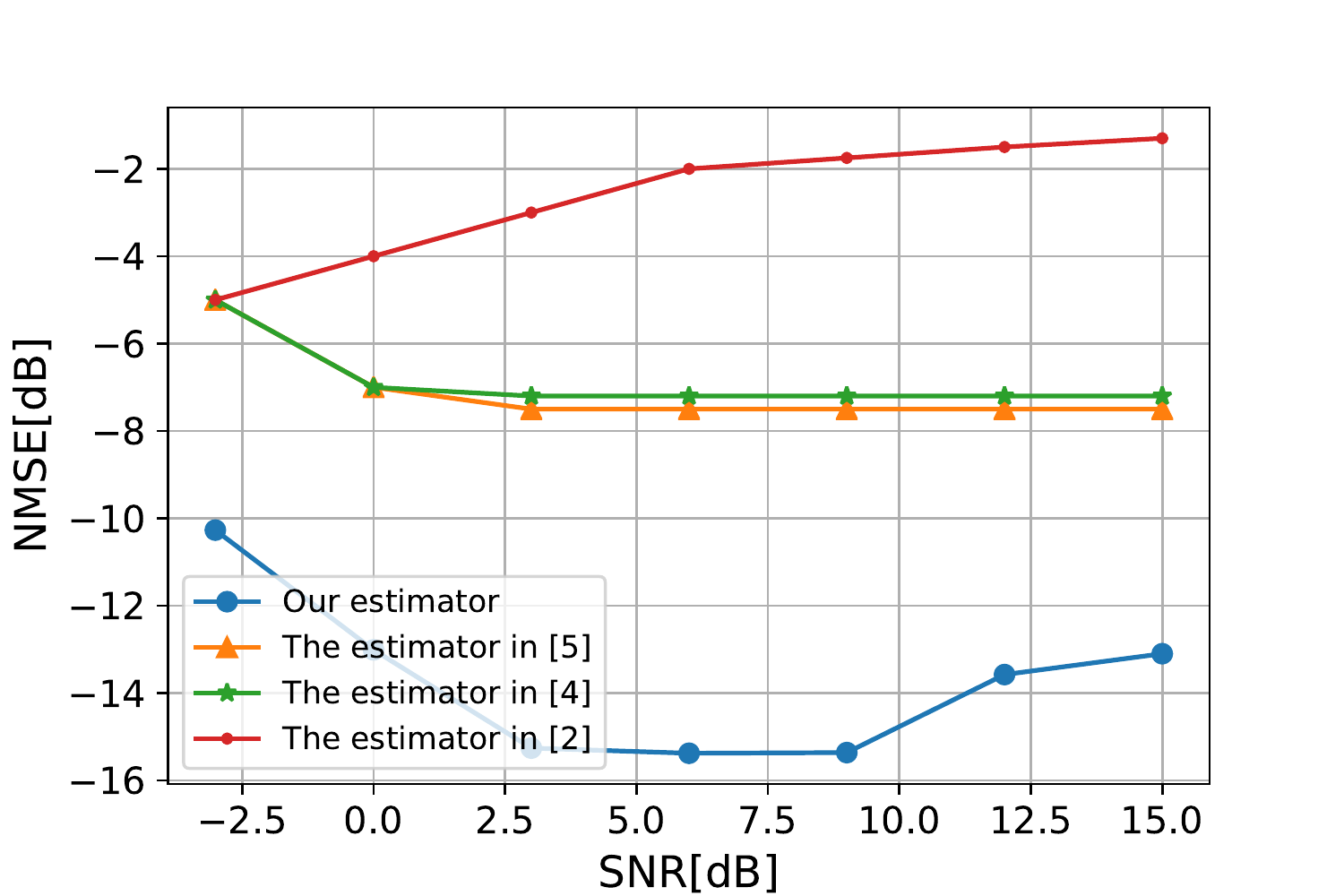}
\caption{The performance comparison of the proposed estimator with state-of-the-art one-bit massive MIMO channel estimators.} \label{fig:massive_mimo_est}
\end{figure}

The performance of the proposed channel estimator is evaluated for the following scenario. We assume that there are $4$ single antenna users and a single base station that has $16$ antennas. Indeed, the number of users are not important as long as $K\ll16$. Users transmit $20$ orthogonal OFDM pilot symbols, each of which has $64$ subcarriers, over a realistic ``extended a pedestrian'' channel model employed in LTE. It is also assumed that pilots are sent at the beginning of each coherence time interval. With these settings, the normalized mean square error (NMSE) of the proposed channel estimator is compared with the other methods that were proposed recently in Fig. \ref{fig:massive_mimo_est}. As can be seen, our deep learning-based channel estimator provides 5-10 dB gain. 

One of the main concerns in deep learning models is the overall computational complexity. For our model, the computational complexity of the first stage is composed of training the $M$ neural networks and generating random samples from the trained neural networks. The former leads to the complexity of $M\mathcal{O}(W^2)$ where $W=32N_f^2$ is the total number of adaptive parameters in the neural network, which stems from the backpropagation algorithm. The latter phase has relatively less complexity, in particular its complexity comes from matrix-vector multiplication. Hence, the first model for channel estimation has a complexity of $M\mathcal{O}(W^2)$. The computational complexity of the second stage is much less than the first stage, because there is a convolutional deep neural network architecture that yields a small number of parameters. In particular, the number of parameters in the unsupervised learning model is on the order of hundreds or thousands \cite{BalDosAndMassiveMIMO}. Hence, the computational complexity of the proposed pipeline is dominated by the supervised learning tasks.

\section{Conclusions} \label{Conc}
In this paper we proposed a novel deep learning-based channel estimator for one-bit massive MIMO in uplink communication. Due to the significant information loss in one-bit quantization, generative models are leveraged as deep learning architectures so as to exploit some prior information in estimating the channel. Our results show that considerable progress can be made with deep generative models for challenging channel estimation problems. As a future work, it can be interesting to consider different types of deep generative models for downlink massive MIMO channel estimation. This is much more challenging than uplink, because many more pilots are required in the downlink channel estimation.

\end{document}